\pdfoutput=1
\documentclass[reprint,aip,apl, twocolumn,floatfix,superscriptaddress]{revtex4-1}
\usepackage{graphicx}
\usepackage{dcolumn}
\usepackage{bm}
\usepackage{amsmath}
\usepackage{lineno}
\usepackage[english]{babel}
\usepackage[utf8]{inputenc}
\usepackage{color}
\usepackage{siunitx}
\usepackage{amsmath}
\usepackage{textcomp, gensymb}
\usepackage{upgreek}
\usepackage{hyperref}
\hypersetup{
    colorlinks=true,
    citecolor=magenta,
    linkcolor=blue,
    filecolor=magenta,      
    urlcolor=magenta,
    }

\begin{document}

\title{Germanium wafers for strained quantum wells with low disorder}

\author{Lucas E. A. Stehouwer}
\affiliation{QuTech and Kavli Institute of Nanoscience, Delft University of Technology, Lorentzweg 1, 2628 CJ Delft, Netherlands}
\author{Alberto Tosato}
\affiliation{QuTech and Kavli Institute of Nanoscience, Delft University of Technology, Lorentzweg 1, 2628 CJ Delft, Netherlands}
\author{Davide Degli Esposti}
\affiliation{QuTech and Kavli Institute of Nanoscience, Delft University of Technology, Lorentzweg 1, 2628 CJ Delft, Netherlands}
\author{Davide Costa}
\affiliation{QuTech and Kavli Institute of Nanoscience, Delft University of Technology, Lorentzweg 1, 2628 CJ Delft, Netherlands}
\author{Menno Veldhorst}
\affiliation{QuTech and Kavli Institute of Nanoscience, Delft University of Technology, Lorentzweg 1, 2628 CJ Delft, Netherlands}
\author{Amir Sammak}
\affiliation{QuTech and Netherlands Organisation for Applied Scientific Research (TNO), Stieltjesweg 1, 2628 CK Delft, The Netherlands}
\author{Giordano Scappucci}
\email{g.scappucci@tudelft.nl}
\affiliation{QuTech and Kavli Institute of Nanoscience, Delft University of Technology, Lorentzweg 1, 2628 CJ Delft, Netherlands}

\date{\today}

\begin{abstract}
We grow strained Ge/SiGe heterostructures by reduced-pressure chemical vapor deposition on 100~mm Ge wafers. The use of Ge wafers as substrates for epitaxy enables high-quality Ge-rich SiGe strain-relaxed buffers with a threading dislocation density of $(6\pm1) \times 10^5~\mathrm{cm}^{-2}$, nearly an order of magnitude improvement compared to control strain-relaxed buffers on Si wafers. The associated reduction in short-range scattering allows for a drastic improvement of the disorder properties of the two-dimensional hole gas, measured in several Ge/SiGe heterostructure field-effect transistors. We measure an average low percolation density of $(1.22\pm0.03)\times10^{10}~\mathrm{cm}^{-2}$, and an average maximum mobility of $(3.4\pm0.1)\times 10^{6}~ \mathrm{cm}^2/\mathrm{Vs}$ and quantum mobility of $(8.4\pm0.5)\times 10^{4}~ \mathrm{cm}^2/\mathrm{Vs}$ when the hole density in the quantum well is saturated to $(1.65\pm0.02)\times10^{11}~\mathrm{cm}^{-2}$. We anticipate immediate application of these heterostructures for next-generation, higher-performance Ge spin-qubits and their integration into larger quantum processors.

\end{abstract}

\maketitle

Strained germanium quantum wells in silicon-germanium heterostructures (Ge/SiGe) have become the leading platform for quantum computation with hole spin qubits.\cite{scappucci_germanium_2021} Single-hole spin qubits and singlet-triplet qubits can be universally controlled,\cite{hendrickx_single-hole_2020, jirovec_singlet-triplet_2021,wang_probing_2022} four-qubit logic has been executed,\cite{hendrickx_four-qubit_2021} and quantum dot systems have been scaled to crossbar arrays comprising 16 quantum dots.\cite{borsoi_shared_2022} Furthermore, the demonstration of a hard superconducting gap in Ge\cite{tosato_hard_2023} motivates the pursuit of coherent coupling of high fidelity Ge spin qubits using crossed Andreev reflection for achieving two-qubit gates over micrometer distances.\cite{choi_spin-dependent_2000,leijnse_coupling_2013} While single-spin qubits have been operated with fidelity as high as 99.99\%,\cite{Lawrie2021SimultaneousThreshold} and rudimentary error correction circuits have been executed,\cite{van_riggelen_phase_2022} quantum coherence limits the operation of larger systems. Although Ge can be isotopically enriched to remove dephasing due to hyperfine interaction,\cite{itoh_high_1993} which can also be achieved by strong confinement,\cite{bosco_fully_2021} hole spin qubits are highly sensitive to charge noise, strain fluctuations, and other types of disorder that can affect the spin-orbit interactions.\cite{bosco_hole_2021,malkoc_charge-noise-induced_2022,terrazos_theory_2021,abadillo-uriel_hole_2022} In addition to optimizing the semiconductor-dielectric interface in qubit devices, further improving the crystalline quality of strained quantum wells\cite{paquelet_wuetz_reducing_2023} appears as a key step to obtain a quieter environment for Ge quantum dots. 

In the absence of suitable SiGe wafers for high-quality and uniform epitaxy, strained Ge quantum wells are commonly deposited on $\mathrm{Si}_{1-x}\mathrm{Ge}_{x}$ strain-relaxed buffers (SRBs) with high Ge composition ($x \approx 0.7-0.8$).\cite{scappucci_germanium_2021} Starting epitaxy from a Si wafer, Ge-rich SiGe SRBs are obtained by composition grading either in a forward-graded process\cite{isella_low-energy_2004} or in a reverse-graded process after the deposition of a thick strain-relaxed Ge layer.\cite{shah_reverse_2008,sammak_shallow_2019} In both cases, the large lattice mismatch between the Si substrate and the Ge-rich SiGe SRB causes a dense misfit dislocation network, with associated threading dislocations that propagate through the quantum well. Moreover, such misfit dislocation network drives significant local strain fluctuations inside the Ge quantum well,\cite{corley-wiciak_nanoscale_2023}, thus challenging the scalability of semiconductor qubits. In Ge/SiGe heterostructures used to host qubits, the threading dislocation density (TDD) is in the range $\approx 10^{6}- 10^{7}$~cm$^{-2}$.\cite{sammak_shallow_2019,jirovec_singlet-triplet_2021} It is not surprising that Si/SiGe heterostructures have smaller TDD ($\approx$$10^{5}$~cm$^{-2}$) because the Si-rich SiGe SRBs have less lattice mismatch to the Si substrate due to the smaller Ge composition ($x \approx 0.2-0.3$). In this Letter we depart from Si wafers and investigate Ge/SiGe heterostructures grown directly on Ge wafers, mitigating the complication of a large lattice-mismatch between Ge-rich SiGe and Si wafers. As a result, we show a significant enhancement of the crystal quality of the heterostructure, as well as a drastic improvement in the disorder properties of the two-dimensional hole gas (2DHG) that it supports.

\begin{figure}
    \centering
	\includegraphics[width=86mm]{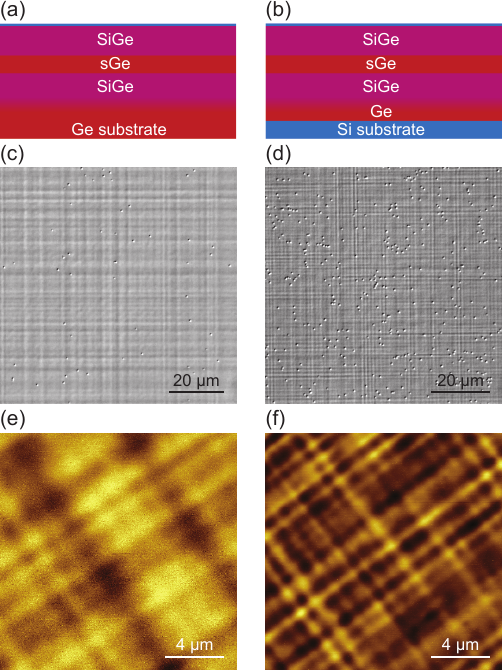}
	\caption{(a) Schematic of a Ge/SiGe heterostructure on a Ge wafer and (b) on a Si wafer. The strained Ge (sGe) quantum wells are grown with the same lattice parameter to SiGe strain-relaxed buffers (SRB). (c) and (d) Comparative optical microscope images of the heterostructures in (a) and (b) after threading dislocation decoration. The images are aligned to the ⟨110⟩ crystallographic axes. (e) and (f) Comparative atomic force microscopy images of the heterostructures in (a) and (b). The images were taken with an alignemnt of about 45 degrees to the ⟨110⟩ crystallographic axes.}
\label{fig:one}
\end{figure}

Schematics in Figs.~\ref{fig:one}(a) and (b) compare heterostructures on a Ge wafer with our control reverse-graded heterostructures on a Si wafer,\cite{lodari_low_2021} the same that supported a four qubit quantum processor and a 16 quantum dot crossbar array.\cite{hendrickx_four-qubit_2021,borsoi_shared_2022} The 100~mm Ge wafers are prepared for epitaxy by an \textit{ex-situ} HF-dip etch followed by \textit{in-situ} bake at $800~\degree \mathrm{C}$. The heterostructure is grown in a high-throughput reduced-pressure chemical vapor deposition tool from high-purity germane and dichlorosilane. The SiGe SRB is $\sim$$2.5~\upmu\mathrm{m}$ thick and obtained by forward step grading of the Si content ($1-x = 0.07, 0.13, 0.17$). 
This approach mirrors the common approach in Si/SiGe heterostructures where the Ge content is forward-graded starting from a Si wafer. Like our previous heterostructures,\cite{sammak_shallow_2019} the SiGe SRB is deposited at $800~\degree \mathrm{C}$ and the growth temperature is reduced to $500~\degree \mathrm{C}$ for the final 200~nm of SiGe below the quantum well and for all the layers above to achieve sharp quantum well interfaces. Importantly, by growing on Ge wafers we avoid the overtensile strain arising from the difference in the thermal expansion coefficients between Ge epilayers and Si substrates.\cite{shah_reverse_2008} Consequently, to achieve 
an in-plane strain ($\epsilon$) in the Ge quantum well similar to our previous heterostructures,\cite{sammak_shallow_2019,lodari_low_2021} here we increase the final Ge content $x$ in the $\mathrm{Si}_{1-x}\mathrm{Ge}_{x}$ SRB to 0.83 (the supplementary material) compared to 0.8 in Refs.~\cite{sammak_shallow_2019,lodari_low_2021}). The thickness of the Ge quantum well (16~nm) and of the SiGe barrier on top (55~nm) are nominally the same compared to Ref.~\cite{lodari_low_2021} for a meaningful comparison of the electrical transport properties.

\begin{figure}[t]
	\includegraphics[width=86mm]{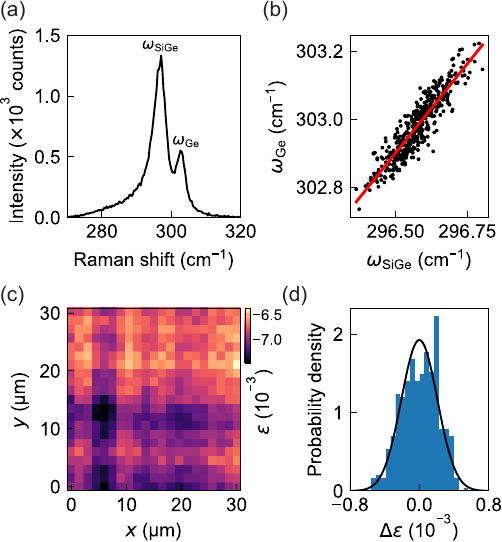}%
	\caption{(a) Typical intensity spectra as a function of the Raman shift for a Ge/SiGe heterostructure on a Ge wafer. The positions of the Raman peaks from the Ge-Ge vibration modes in the strained Ge quantum well and in the SiGe layer are marked as $\omega_{\mathrm{Ge}}$ and $\omega_{\mathrm{SiGe}}$, respectively. (b) Distribution of Raman peak positions of the Ge-Ge modes obtained by analyzing Raman spectra over an area of $30\times30~\mathrm{\upmu m^2}$ and linear fit (red). (c) Raman strain map,  corresponding to the $\omega_{\mathrm{Ge}}$ Raman shifts in (b). The map is aligned to the ⟨110⟩ crystallographic axes. (d) Strain fluctuations from the Raman map in (c) and normal distribution fit (black). Counts are normalized such that the area under the curve integrates to one.}
\label{fig:two}
\end{figure}

Figure~\ref{fig:one}(c) and (d) show comparative images by Nomarski microscopy of the heterostructures on a Ge and on a Si wafer after decorating the threading dislocations by \textit{in-situ} HCl vapor etching.\cite{bogumilowicz_chemical_2005} We quantify the TDD by counting the number of decorated threading dislocations from multiple images taken across the wafer. Changing substrate from Si to Ge improves the TDD almost an order of magnitude, from $(5.3\pm0.3) \times 10^6~\mathrm{cm}^{-2}$ to $(6\pm1) \times 10^5~\mathrm{cm}^{-2}$. Consequently the average TD separation ($1/\sqrt{\mathrm{TDD}}$) becomes much longer, from $\sim4.3~\upmu\mathrm{m}$ to $\sim13~\upmu\mathrm{m}$. Comparative atomic-force microscopy images in Figs.~\ref{fig:one}(e) and (f) show the typical cross hatch pattern arising from the strain-releasing misfit dislocation network within the SiGe SRB on Ge and Si wafers, respectively. The root mean square surface roughness of both heterostructures is similar at $\sim$1.5~nm. However, the heterostructure grown on a Ge wafer shows cross-hatch undulations with a longer wavelength and weaker high-frequency components of the Fourier spectrum (the supplementary material). This observation supports the intuition that the Ge-rich SiGe SRB has a less dense network of misfit dislocations when grown on a Ge wafer, as the lattice mismatch with the substrate is smaller compared to when it is grown on a Si wafer.

We further characterize the heterostructure on the Ge wafer by scanning Raman spectroscopy over an area of $30\times30 ~\upmu\mathrm{m^2}$, much larger than the length scale of the cross-hatch pattern features. In particular, we determined the in-plane strain in the quantum well $\epsilon$ and analyzed the origin and bandwidth of its fluctuations. The representative spectrum in Fig.~\ref{fig:two}(a) was obtained with a 633~nm red laser and shows two clear Raman peaks originating from the Ge-Ge vibration modes in the strained Ge quantum well ($\omega_{\mathrm{Ge}}$) and in the SiGe layer ($\omega_{\mathrm{SiGe}}$). The distribution of these Raman shifts in Fig.~\ref{fig:two}(b) shows a strong correlation, with a slope $\Delta\omega_{\mathrm{Ge}}/\Delta\omega_{\mathrm{SiGe}}=1.05\pm0.02$. Comparing to predictions by Eq.~5 in Ref.~\cite{kutsukake_origin_2004}, we argue that the distribution of the Raman shift in the Ge quantum well is mainly driven by strain fluctuations in the SiGe SRB (expected $\Delta\omega_{\mathrm{Ge}}/\Delta\omega_{\mathrm{SiGe}}\sim0.83$), rather than compositional fluctuations ($\Delta\omega_{\mathrm{Ge}}/\Delta\omega_{\mathrm{SiGe}}\sim 0.25$). Figure~\ref{fig:two}(c) shows the Raman strain map of the Ge quantum well calculated using  $\epsilon = (\omega_{\mathrm{Ge}} - \omega_0) / b^{\mathrm{Ge}}$, where $\omega_0 = 299.9~\mathrm{cm}^{-1}$ is the Raman shift for bulk Ge and $b^{\mathrm{Ge}}=-440~\mathrm{cm}^{-1}$ is the Ge phonon strain shift coefficient.\cite{pezzoli_phonon_2008} We identify signatures of the cross-hatch pattern, with regions featuring higher and lower strain around a mean strain value $\overline{\epsilon}= (-6.9\pm0.2)\times10^{-3}$. This is similar to the compressive strain measured in our Ge quantum wells on a Si wafer,\cite{sammak_shallow_2019} validating our heterostructure design and comparative analysis. The statistics of the lateral strain map are shown in 
Fig.~\ref{fig:two}d. The strain fluctuations $\Delta\epsilon$ around the average $\overline{\epsilon}$ follow a normal distribution with a standard deviation of $2\times10^{-4}$. The bandwidth of the strain fluctuations is reduced when compared to the strain fluctuations from the heterostructure on a Si wafer,\cite{corley-wiciak_nanoscale_2023} pointing to a more uniform strain landscape.

\begin{figure}
	\includegraphics[width=86mm]{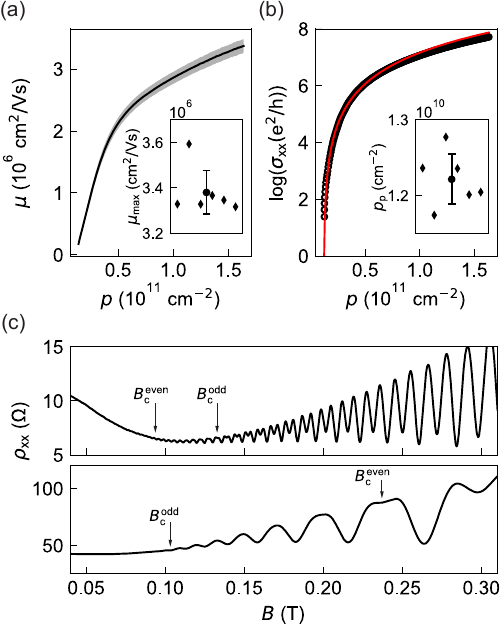}%
	\caption{(a) Mobility $\mu$ mean (blue) and standard deviation (shaded) as a function of density $p$ obtained from measurements at $T=70~\mathrm{mK}$ of six Hall bar devices from the same wafer. The inset shows the maximum mobility $\mu_\mathrm{max}$ from all the devices and average value $\pm$ standard deviation (black). (b) Conductivity $\sigma_{\mathrm{xx}}$ as a function of $p$ (circles) for one device and fit to the percolation theory in the low density regime (solid red line). The inset shows the percolation density $p_\mathrm{p}$ from all the devices and average value $\pm$ standard deviation (black). (c) Longitudinal resistivity $\rho_{\mathrm{xx}}$ as a function of perpendicular magnetic field $B$ measured at a density of $1.5\times10^{11}~\mathrm{cm}^{-2}$ (upper panel) and $6\times10^{10}~\mathrm{cm}^{-2}$ (lower panel). $B_\mathrm{c}^\mathrm{even}$ and $B_\mathrm{c}^\mathrm{odd}$ indicate the critical magnetic fields for resolving even and odd filling factors, corresponding to the cyclotron and the spin gap, respectively.} 
\label{fig:three}
\end{figure}

The structural characterization highlights the improvement in crystal quality when growing a Ge-rich SRB on a Ge wafer instead of a Si wafer. Next, we show how a better and more uniform crystalline environment improves the disorder properties of the 2DHG. 
We fabricate six Hall-bar shaped heterostructure field effect transistors (H-FETs) on a $2\times2$~cm$^{2}$ coupon from the center of the 100~mm wafer with a similar process as in Ref.~\cite{lodari_low_2021} 
We accumulate a 2DHG inside the Ge quantum well by applying a negative DC gate voltage ($V_{\mathrm{g}}$) to the top gate of the H-FETs and we increase the density $p$ in the 2DHG above the percolation density ($p_{\mathrm{p}}$) by making $V_{\mathrm{g}}$ more negative. We use standard four-probe low-frequency lock-in techniques for mobility-density and magnetotransport characterization of all devices in a dilution refrigerator equipped with a cryo-multiplexer\cite{paquelet_wuetz_multiplexed_2020} ($T = 70$~mK measured at the mixing chamber).

Figure~\ref{fig:three}(a) shows the density-dependent mobility curve (solid line), obtained by averaging over the six H-FETs, together with the standard deviation (shaded region). We observe a tight distribution over the entire density range, indicating a very uniform disorder potential landscape. The mobility increases steeply with increasing density, due to increasing screening of the remote impurity charges, most likely at the semiconductor-dielectric interface. At higher densities ($p > 5\times10^{10}~\mathrm{cm}^{-2}$), the mobility increases less rapidly, signaling the relevance of scattering from impurities within or in the proximity of the quantum well.\cite{monroe_comparison_1993} We observe a maximum mobility $\mu_\mathrm{max}$ in the range of $3.3-3.6\times10^{6}~\mathrm{cm^{2}/Vs}$ over the six investigated H-FETS (Fig.~\ref{fig:three}(a), inset), from which we extract an average $\mu_\mathrm{max}=(3.4\pm0.1)\times 10^{6}~ \mathrm{cm}^2/\mathrm{Vs}$ at a saturation density $p_\mathrm{sat}=(1.65\pm0.02)\times10^{11}\mathrm{cm}^{-2}$, corresponding to a long mean free path of $23~\mathrm{\upmu m}$. Figure~\ref{fig:three}b shows the longitudinal conductivity $\sigma_\mathrm{xx}$ as a function of density $p$ for a representative H-FET. We extract the percolation density $p_\mathrm{p}$ from fitting to percolation theory,\cite{tracy_observation_2009} $\sigma_\mathrm{xx} \propto (p-p_\mathrm{p})^{1.31}$. The inset shows $p_\mathrm{p}$ for the six H-FETs, from which we extract an average percolation density $p_\mathrm{p}=(1.22\pm0.03)\times10^{10}~\mathrm{cm}^{-2}$.

Compared to our control heterostructures on a Si wafer supporting qubits,\cite{lodari_low_2021,hendrickx_four-qubit_2021} the maximum mobility is more than 15 times larger and the percolation density is more than 1.5 times smaller. We speculate that this significant improvement, throughout the whole density range, is associated with the suppression of short-range scattering from dislocations within the quantum well. Furthermore, the mobility has not yet saturated indicating that it is still limited by long-range scattering from impurities at the dielectric interface, leaving room for further improvement. In fact, our maximum mobility, reproducible across multiple devices, is less than the value $\mu=4.3\times 10^{6}~ \mathrm{cm}^2/\mathrm{Vs}$ measured on a single H-FET in Ge/SiGe grown on a Si wafer,\cite{myronov_holes_2023} likely because the dielectric interface in our samples is much closer to the channel (55~nm compared to 100~nm in Ref.~\cite{myronov_holes_2023}). 

The low level of disorder is confirmed by high-quality magnetotransport characterization observed in all devices. Figure~\ref{fig:three}(c) shows representative magneto-resistivity curves from an H-FET at fixed densities of $1.5\times10^{11}~\mathrm{cm}^{-2}$ (upper panel) and $6\times10^{10}~\mathrm{cm}^{-2}$ (lower panel). The measurements were performed by keeping $V_\mathrm{g}$ constant and sweeping the perpendicular magnetic field $B$. For each longitudinal resistivity ($\rho_\mathrm{xx}$) curve we extract the pair of critical magnetic fields for resolving the cyclotron and the spin gap, $\{B_\mathrm{c}^\mathrm{even},B_\mathrm{c}^\mathrm{odd}\}$, corresponding, respectively, to even and odd filling factors $\nu = h p/ e B$ in the Shubnikov --de Haas oscillations minima. Due to the very small Landau level broadening at high density, the cyclotron gap (see upper panel of Fig.~\ref{fig:three}(c)) is resolved already at very low critical fields and the spin gap only a few oscillations later, $\{B_\mathrm{c}^\mathrm{even},B_\mathrm{c}^\mathrm{odd}\}=\{0.08,0.13\}~\mathrm{T}$. However, at low density the order is reversed and the spin gap is resolved earlier than the cyclotron gap, $\{B_\mathrm{c}^\mathrm{even},B_\mathrm{c}^\mathrm{odd}\}=\{0.24,0.095\}~\mathrm{T}$ (see lower panel of Fig.~\ref{fig:three}(c)). This is typical to 2DHGs in Ge/SiGe and occurs when the spin gap is more than half of the cyclotron gap, due to the increased perpendicular $g$-factors of holes at low density.\cite{lu_density-controlled_2017,lodari_light_2019,myronov_holes_2023}

The combination of these two aspects, the very low level of disorder and the increasingly large spin gap at low density, makes the canonical methods\cite{coleridge_small-angle_1991,coleridge_effective_1996} for extracting the effective mass $m^*$ and single-particle lifetime $\tau_\mathrm{q}$ not straightforward throughout the investigated density range and will be pursued in a further study. However, we may still estimate the quantum mobility $\mu_\mathrm{q}=e\tau_\mathrm{q}/m^*$ without making assumptions on  $m^*$ and $\tau_\mathrm{q}$ using the expression $\mu_\mathrm{q}=(1+\sqrt{B_\mathrm{c}^\mathrm{odd}/B_\mathrm{c}^\mathrm{even}})/2B_\mathrm{c}^\mathrm{odd}$ (the supplementary material). With this analysis, we obtain a maximum quantum mobility in the range of $7.7-9.1\times10^{4}~\mathrm{cm^{2}/Vs}$ over the six investigated H-FETS (the supplemental material), from which we extract an average maximum 
$\mu_\mathrm{q}=(8.4\pm0.5)\times 10^{4}~ \mathrm{cm}^2/\mathrm{Vs}$ at saturation density. This value should be considered as a conservative estimate of $\mu_\mathrm{q}$, as the onset of Shubnikov --de Haas oscillations in our high-quality samples might be limited by small density inhomogeneities at low magnetic field.\cite{qian_quantum_2017} The maximum $\mu_\mathrm{q}$ is over three times larger than that of our control heterostructures on a Si wafer,\cite{lodari_low_2021,hendrickx_four-qubit_2021} and approximately two times larger compared to the heterostructures on a Si wafer in Ref.~\cite{myronov_holes_2023} These results highlight the significantly improved short-range scattering in 2DHGs when the Ge-rich SiGe SRB is grown on a Ge substrate, setting a benchmark for holes in group IV semiconductors.

In summary, we challenged the mainstream approach to deposit Ge/SiGe heterostructures on Si wafers and instead, we started epitaxy on a Ge wafer. We demonstrate a more uniform crystalline environment with fewer dislocations and in-plane strain fluctuations compared to control heterostructures supporting a four-qubit quantum processors. Future investigations using X-ray diffraction spectroscopy to map the complete strain tensor\cite{corley-wiciak_nanoscale_2023} could provide insights into the local strain modifications and fluctuation caused by nanostructured metallic gates. The disorder properties of the 2DHG are also greatly improved, with reproducible ultra-high mobility, very low percolation density, and high quantum mobility. Considering these heterostructures on Ge wafers as a proof of principle, the electrical transport metrics are likely to further improve by routine optimization of the heterostructure design and chemical vapor deposition process. We anticipate immediate benefit of using these heterostructures for improved uniformity and yield in large quantum dot arrays. Future studies on charge noise and qubit performance may also provide insight in possible improved quantum coherence. Furthermore, it would be of significant interest to explore whether suppressing the dislocation network in the SiGe SRB could improve the performance of superconducting microwave resonators integrated atop the heterostructure, towards the development of hybrid superconductor-semiconductor architectures.

\vspace{\baselineskip}
See the supplementary material for secondary ions mass spectroscopy, Fourier transform of the atomic force microscopy images, and the derivation of the expression for quantum mobility.

\vspace{\baselineskip}
This work was supported by the Netherlands Organisation for Scientific Research (NWO/OCW), as part of the Frontiers of Nanoscience program. A.T. acknowledges support through a projectruimte associated with the Netherlands Organization of Scientific Research (NWO). This research was supported by the European Union’s Horizon 2020 research and innovation programme under the grant agreement No. 951852.

\section*{Author Declarations}
The authors have no conflict to disclose

\section*{Data availability}
The data sets supporting the findings of this study are openly available in 4TU Research Data at \url{https://doi.org/10.4121/52a5bcf9-8e15-4871-b256-10f90e320ecc}, Ref~\cite{dataset}.

\end{document}


\title{Supplementary material: Germanium wafers for strained quantum wells with low disorder}

\author{Lucas E. A. Stehouwer}
\affiliation{QuTech and Kavli Institute of Nanoscience, Delft University of Technology, Lorentzweg 1, 2628 CJ Delft, Netherlands}
\author{Alberto Tosato}
\affiliation{QuTech and Kavli Institute of Nanoscience, Delft University of Technology, Lorentzweg 1, 2628 CJ Delft, Netherlands}
\author{Davide Degli Esposti}
\affiliation{QuTech and Kavli Institute of Nanoscience, Delft University of Technology, Lorentzweg 1, 2628 CJ Delft, Netherlands}
\author{Davide Costa}
\affiliation{QuTech and Kavli Institute of Nanoscience, Delft University of Technology, Lorentzweg 1, 2628 CJ Delft, Netherlands}
\author{Menno Veldhorst}
\affiliation{QuTech and Kavli Institute of Nanoscience, Delft University of Technology, Lorentzweg 1, 2628 CJ Delft, Netherlands}
\author{Amir Sammak}
\affiliation{QuTech and Netherlands Organisation for Applied Scientific Research (TNO), Stieltjesweg 1, 2628 CK Delft, The Netherlands}
\author{Giordano Scappucci}
\email{g.scappucci@tudelft.nl}
\affiliation{QuTech and Kavli Institute of Nanoscience, Delft University of Technology, Lorentzweg 1, 2628 CJ Delft, Netherlands}

\date{\today}
\pacs{}

\maketitle

\section{Secondary ions mass spectroscopy}
\begin{figure}[H]
    \centering
	\includegraphics[width=86mm]{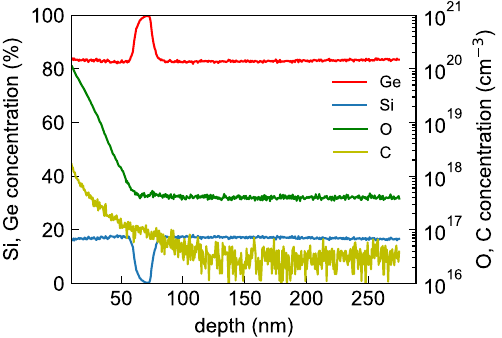}
	\caption{Secondary ions mass spectroscopy of Ge/SiGe
heterostructures on a Ge wafer, showing Si (blue), Ge (red), O (green), C (yellow) concentration depth profiles}  
\label{fig:S1}
\end{figure}

Secondary ion mass spectroscopy (SIMS) was carried out to determine the chemical composition of the Ge/SiGe heterostructure on the Ge substrate. Supplementary Fig.~\ref{fig:S1} overlays the Ge (red), Si (blue), O (green), and C (yellow) signals. The data confirms the chemical composition $x=0.83$ of the Si$_{1-x}$Ge$_x$ buffer and barrier surrounding the Ge quantum well. The oxygen and carbon concentrations far from the surface are below the detection limit.  The rising O and C signals near the surface are routine in SIMS and should be considered an artifact of the measurement.

\section{Atomic force microscopy}
We analysed the cross-hatch patterns shown in Figs.~1(e) and (f) in the main text by performing a 2D Fourier transform using Gwyddion.\cite{necas_gwyddion:_2012} The results are shown in supplementary Figs.~\ref{fig:S2}(a) and (b) for the heterostructure grown on a Ge wafer and Si wafer, respectively. A comparison of the two images shows that the dominant frequencies are spaced closer together in the 2D Fourier transform of the heterostructure on a Ge substrate than those in the heterostructure on the Si substrate. This confirms that using a Ge wafer as a substrate results in a longer wavelength of the cross-hatch pattern compared to when a Si wafer is used, a consequence of the reduced misfit dislocation network present in this heterostructure.  

\begin{figure}[H]
    \centering
	\includegraphics[width=86mm]{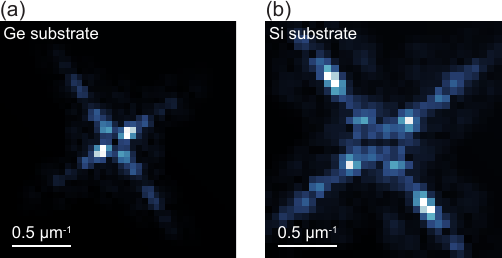}	\caption{(a) 2D Fourier transform of AFM image in Fig.~1(e) in the main text for heterostructures on a Ge wafer. (b) 2D Fourier transform of AFM image in Fig.~1(f) in the main text for heterostructures on a Si wafer.}  
\label{fig:S2}
\end{figure}

\section{Quantum mobility}
In these Ge/SiGe quantum wells on Ge wafers, the combination of the very low level of disorder and the increasingly large spin gap makes an estimate of the effective $g$-factor $g^*$  and effective mass $m^*$ challenging from the analysis of the thermal activation of the resistivity minima in the Shubnikov -de Haas oscillations corresponding to the cyclotron and spin gap. The same applies to the standard analysis\cite{coleridge_small-angle_1991} for extracting the single-particle lifetime $\tau_\mathrm{q}$, which measures the time for which a momentum eigenstate can be defined even in the presence of scattering,\cite{das_sarma_single-particle_1985} and the associated quantum mobility $\mu_\mathrm{q}=e\tau_\mathrm{q}/m^*$. However, just like the classical mobility is typically measured without knowing the effective mass $m^*$ or the scattering time $\tau_\mathrm{t}$, we may also estimate the quantum mobility without a direct measurement of $m^*$ and $\tau_\mathrm{q}$. 
Assuming that Landau levels have a Gaussian broadening $\Gamma=\hbar/2\tau_\mathrm{q}$\cite{das_sarma_single-particle_1985} that increases as $\sqrt{B}$, where $B$ is the perpendicular magnetic field, Refs.~\cite{sammak_shallow_2019,myronov_holes_2023} show that:
\begin{equation}    
g^* = \frac{2m_e}{m^*}\frac{\sqrt{B_\mathrm{c}^\mathrm{even}}}{\sqrt{B_\mathrm{c}^\mathrm{even}}+\sqrt{B_\mathrm{c}^\mathrm{odd}}}    
\label{eq:gfactor}
\end{equation}
where $B_\mathrm{c}^\mathrm{even}$ and $B_\mathrm{c}^\mathrm{odd}$ are the critical magnetic fields for resolving the cyclotron and the spin gap, corresponding, respectively, to even and odd filling factors $\nu = h p/ e B$ in the Shubnikov--de Haas oscillations minima of the magnetoresistivity. Evaluating the level broadening at $B=B_\mathrm{c}^\mathrm{odd}$ yields:
\begin{equation}
    \Gamma=g^*\mu_\mathrm{B}B_\mathrm{c}^\mathrm{odd}
\label{eq:gammaBc}
\end{equation}
where the Bohr magneton $\mu_\mathrm{B}=e\hbar/2m_e$. Expressing the level broadening in terms of quantum mobility yields: 

\begin{equation}
    \Gamma=e\hbar/2m^*\mu_\mathrm{q}.
\label{eq:gammamuq}
\end{equation} 
By inserting $g*$ from Eq.~\ref{eq:gfactor} and $\Gamma$ from Eq.~\ref{eq:gammamuq} into Eq.~\ref{eq:gammaBc}, we obtain an expression for quantum mobility that is independent of $m^*$ and $g*$:  

\begin{equation}
\mu_\mathrm{q} =\frac{1+\sqrt{B_\mathrm{c}^\mathrm{odd}/B_\mathrm{c}^\mathrm{even}}}{2B_\mathrm{c}^\mathrm{odd}}.    
\label{eq:muq}
\end{equation}

We apply this analysis to the magnetoresistivity curves at saturation density for all six heterostructures field effect transistors and show in Supplementary Fig.~\ref{fig:S3} the obtained quantum mobility values and the average with standard deviation (black).

\begin{figure}[H]	
    \centering
	\includegraphics[width=50mm]{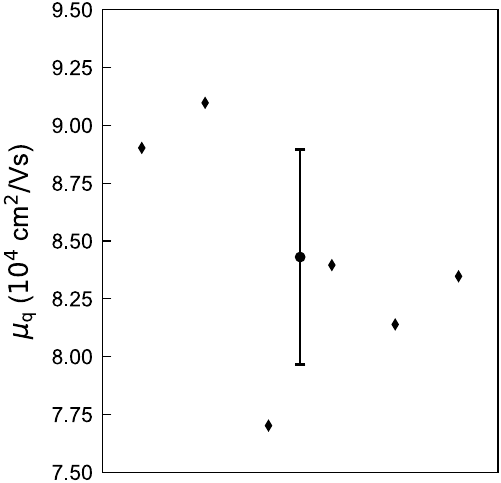}
	\caption{Quantum mobility for Ge/SiGe on Ge substrates measured at saturation density for all the six heterostructure field effect transistors.} 
\label{fig:S3}
\end{figure}